\documentclass[12pt]{article}
\newcommand{\eq}{\begin{equation}}
\newcommand{\en}{\end{equation}}
\newcommand{\eqa}{\begin{eqnarray}}
\newcommand{\ena}{\end{eqnarray}}

\newcommand{\br}{\langle}
\newcommand{\kt}{\rangle}

\newcommand{\wh}{\widehat}

\newcommand{\lan}{\langle}
\newcommand{\ran}{\rangle}
\newcommand{\nonu}{\nonumber}

\newcommand{\dep}{\partial}   

\newcommand{\phir}{{{\phi}}}
\newcommand{\phil}{{{\phi^0}}}

\begin{document}
\begin{titlepage}
\vskip0.5cm
\begin{flushright}
SISSA REF 35/2003/FM\\
gef-th-04/03\\
 SPhT-T03/049\\
 DFTT 11/03\\
\end{flushright}
\vskip0.5cm
\centerline{\Large\bf  A new class of short distance universal }
\vskip 0.2cm
\centerline{\Large\bf  amplitude ratios.}
\vskip 1.3cm
\centerline{M. Caselle$^a$,
P.Grinza$^b$,  R. Guida $^c$ and N. Magnoli$^{d}$}
 \vskip 1.0cm
 \centerline{\sl $^a$  Dip. di Fisica 
Teorica dell'Universit\`a di Torino and I.N.F.N.
 via P.Giuria 1, I-10125 Turin,Italy}
 \vskip 0.2cm
 \centerline{\sl $^b$ SISSA and I.N.F.N, via Beirut 2-4, I-34014 Trieste, 
 Italy}
 \vskip 0.2cm
 \centerline{\sl $^c$ Service de Physique Th\'eorique de Saclay, 
 CEA/DSM/SPhT-CNRS/SPM/URA 2306,}
 \centerline{ \sl    Orme des Merisiers, F-91191 Gif-sur-Yvette Cedex, France}
 \vskip 0.2cm
 \centerline{\sl $^d$  Dipartimento di Fisica,
 Universit\`a di Genova and
 I.N.F.N.,}
 \centerline{\sl via Dodecaneso 33, I-16146 Genova, Italy}
 \vskip .6 cm

\begin{abstract}
We propose a new class of universal amplitude ratios which involve the
first terms of the short distance expansion of the correlators 
of a statistical model in the vicinity of a critical point.
We will describe the critical system with  a conformal field theory (UV fixed point) 
perturbed by an appropriate relevant operator.
In two dimensions the exact knowledge of the UV fixed point allows for 
accurate predictions of the ratios  and in many nontrivial 
integrable perturbations they can even be  evaluated exactly.
In three dimensional $O(N)$ scalar systems feasible extensions of some existing results
should allow to obtain perturbative expansions for the ratios. 
By construction these universal ratios are a perfect tool to explore the short
distance properties of the underlying quantum field theory even in regimes
where the correlation length and one point functions are not accessible
in experiments or simulations.

\end{abstract}
\end{titlepage}

\setcounter{footnote}{0}
\def\thefootnote{\arabic{footnote}}

\section{Introduction}
Universal amplitude ratios play a major role in modern statistical
mechanics~\cite{zinn,ahp91,PVReport} of critical systems.
Like critical indices these ratios can be used  
to characterize and identify universality classes. In this respect they are in
general more useful than the critical indices since they are usually simpler to
measure and can vary by a  larger amount, thus allowing a more efficient
discrimination among different universality classes. 

In this letter we  propose a new class 
of  universal ratios related to the
short distance expansion of the correlators of a statistical  model
near criticality: by construction they provide a  perfect tool to probe
the short distance properties of the underlying quantum field theory (QFT in the following)
even in regimes
where the correlation length and one point functions are not accessible
in experiments or simulations. 

We shall discuss the construction of these ratios with a  general formalism
(for a $d$-dimensional QFT with a generic operator content), 
but   we will be mainly interested  to two dimensional systems 
where the necessary assumptions are  satisfied in many nontrivial cases.  
In particular  we have in mind some   integrable perturbations of Conformal Field Theories (CFT)
in which case these short distance universal ratios can be evaluated \textit{exactly}.

This letter is organized as follows. In sect.2, after a short reminder of known
results about QFT and standard amplitude ratios, we shall construct
our new universal ratios. Then in sect.3 we shall discuss,
as a topical example, the case of the magnetic perturbation of the Ising model. 

\section{Universal Ratios}
Let us consider a statistical model near criticality, 
i.e. a $d-$dimensional QFT with operators $\phil_\alpha $, regularized by a  length cut-off 
$a\ll \xi$ ($\xi$ being the correlation length of the system).
We will assume that  properties of 
the statistical model  in the the short distance limit $a\ll r\ll \xi$ are described 
by a renormalized QFT obtained by an  appropriate relevant perturbation $\int d^dx\; g_i \phir_i$
of a CFT. 
(We denote generic renormalized operators  by $\phir_\alpha$, 
and their scaling dimension by  $ 2 \Delta_\alpha$; 
latin indices $i,j\cdots$ will be reserved to relevant operators, such that $ 2 \Delta_j <d$.
As a consequence the coupling $g_i$ of the perturbing operator $\phir_i$ 
will have   scaling dimension $d-2 \Delta_i$,  such $\xi^{-1}\propto g_i^{1/(d-2 \Delta_i)}$.)
In general any regularized operator will be equal 
to an appropriate mixing of 
renormalized operators with  equal or lower scaling dimension, plus irrelevant corrections.
The mixing can happen only if appropriate selection rules 
on scaling dimensions of operators and coupling  
are satisfied and symmetries of the theory are preserved. 
\textit{We assume that  no mixing occurs among the relevant operators of the QFT}
 to which we  restrict 
our analysis. 
In this case we are left with simple multiplicative relations
 (up to irrelevant corrections):
$
 \phil_j= Z_j^{-1} \phir_j +\cdots
$
in which the factors $Z_j$ are the only non-universal avatar of the original  statistical model.
The invariance of the perturbation, 
$\int g_i^0 \phil_i=\int g_i \phir_i+\cdots$,
 implies the relation $g_i^0=Z_i g_i$.
It is clear that any  scale invariant constant built up from relevant operators and coupling
that is invariant by the rescaling 
\begin{eqnarray}\label{metricScaling}
\phi_j \rightarrow K_j^{-1} \phi_j; \;\;\; g_i \rightarrow K_i g_i
\end{eqnarray}
will not depend on $Z_j$ and as a consequence  will be universal.
(We borrow from  ~\cite{Fioravanti:2000xz} the name ``metric factors'' 
for our  normalization constants $K_j$.)

Let us now  review some known  universal amplitude ratios related to VEV and correlators
(For a thorough review and additional references see~\cite{ahp91,PVReport}.)

Let us start with ratios involving one point functions of relevant operators
and denote  
\eq \label{vevji}
\lan \phi_j \ran_i= A_j^{i} \; g_i^{\frac{2 \Delta_j}{d-2 \Delta_i} }
\en 
the vacuum expectation value (VEV) of the operator $\phi_j$ in the theory obtained
perturbing with respect to $\phi_i$.
Clearly, the scaling  Eq.(\ref{metricScaling}) is equivalent to replace 
\eq
A_j^{i}\rightarrow K_j A_j^{i} (K_i)^{\frac{2 \Delta_j}{d-2 \Delta_i} }
\en 
in  Eq.(\ref{vevji}).
The only remaining task to obtain universal ratios is to
construct suitable combinations of the VEV (or of their derivatives wrt $g_i$) 
so as to eliminate all the metric 
factors (see \cite{Lassig:1991xy,Delfino:2003zs,Mussardo:2000qg,Fioravanti:2000xz} for  recent applications). 
An exhaustive list of these
combinations can be
found, for instance, in~\cite{ahp91,PVReport}.

A similar approach can be followed also if we take, instead of one point
functions, the expectation value of two (or more points) functions.
Much work has been done in the past years in the context of $O(N)$ scalar models:
for instance it was soon realized (see~\cite{fisher,PVReport} and references therein)
that the connected spin spin correlator as a function of   $r/\xi$ is a 
universal function (up to an overall constant) 
whose asymptotics can be parametrized in terms of universal constants.

When studying connected correlators in the regime $r \gg \xi\gg a$
one has access to universal
ratios of the so
called ``overlap amplitudes''. 
This line of research have been  recently pushed forward
in the case of the 2d Ising model perturbed by a magnetic field ~\cite{ch00}
and for the thermal perturbation of the 3d Ising model~\cite{chp00}.

The complementary possibility 
is to look at the short distance behavior $a\ll r \ll \xi$ of correlators.
In the context of $O(N)$ scalar models 
Wilson's Operator Product Expansion (OPE) revealed to be~\cite{brezinn}
the appropriate theoretical  tool to deal with short distance regime
and $\epsilon=4-d$ expansion techniques have been applied
 to describe correlators' asymptotics
and to compare them with simulations or with experiments (see~\cite{PVReport}).

In this short letter we  propose a new 
class of universal amplitudes --obtained from ratios of correlators in short distance regime--
that do not require additional measures of $\xi$ or VEV: the idea is that
if one is looking at very small perturbations of the critical point 
it might happen that  
the correlation length 
and VEV estimates are not accessible in the particular experimental or numerical
 realization in which one is interested.

We pursue to describe short distance asymptotics 
by applying a specialization of the  OPE technique to perturbed CFT recently proposed in~\cite{zayl,mz,gm1,gm2}
and known as InfraRed Safe approach (IRS).
Roughly speaking, the main ingredient of the IRS approach 
is simply to choose  a renormalization scheme in which
the Wilson's coefficients of the OPE have a regular, IR safe, perturbative expansion with
respect to the coupling and then to expand them  in OPE.
(See ~\cite{gm1,gm2} for all-orders IRS formulas and 
many cautionary remarks on their range of validity.)
The approach can be extended in principle to $n-$points  correlators and  irrelevant  operators, but in
the following we shall concentrate on the two point functions of relevant operators only,
because the hypotheses of non mixing and of absence of divergences --see below--
are likely to be satisfied in this case and because in more complicate generalizations
the amplitude ratios will be  too difficult to study from a numerical point
of view.

 We choose as expansion 
variable the scaling combination $t \equiv g_i r^{d-2 \Delta_i}\propto (r/\xi)^{d-2 \Delta_i}$ 
and rescale
the correlators
as follows
 \eqa
F_{jk}^{i}(t) \equiv r^{2 \Delta_j + 2 \Delta_k} \lan \phi_j \phi_k \ran_i .
\ena
(Notice that $\lan \phi_j \phi_k \ran_i $
 is just the expectation value of the two operators
 and 
\textit{ not} the connected correlator.)
In this notations IRS gives 
 \eqa
&&
F_{jk}^{i}(t)\sim\sum_{\alpha,n} A_\alpha^{i} \wh{\frac{
\partial^n C_{jk}^\alpha}{\partial g_i^n}} 
\frac{{t}^{\frac{2\Delta_\alpha }{d-2 \Delta_i}+n}}{n!} 
              \equiv\sum_a f_{j k a}^{i} t^{z_{j k a}^{i}}
\label{eq1}
\ena
 where the constants $f_{j k a}^{i}$ are suitable combinations of  
 $\wh{{\partial^n C_{jk}^\alpha}/{\partial g_i^n}}$
(derivatives of Wilson's
 coefficients evaluated at $g_i=0$ and scaled by appropriate powers of $r$ to obtain adimensional
quantities)
 and of the amplitudes $A_{\alpha}^{i}$ of operators' VEV, defined as in  Eq.(\ref{vevji});
  $z_{j k a}^{i}$ are appropriate combinations of scaling dimensions.
 Notice that in Eq.(\ref{eq1}) 
\textit{we assumed a powerlike short-distance expansion in terms of the variable $t$} 
(at least up to the order we will need): 
in general it may happen  that the underlying QFT contains pairs of
``resonant'' operators and terms proportional to powers of  $\log t$ may appear in the
correlators at some order in the expansion 
(the typical example of this situation is the thermal perturbation
of the 2d Ising model). If logs appear in the terms of the expansion 
the present treatment cannot be applied as it is (we shall deal with these peculiar cases in
forthcoming publications).

 The key  point for the construction of universal ratios is that
the scaling Eq.(\ref{metricScaling})
is equivalent to the nontrivial  rescaling 
\eq
f_{j k a}^{i}\rightarrow K_j K_k K_i^{z_{j k a}^{i}} f_{j k a}^{i}
\en
(mixing factors from $\br \phir_\alpha\kt$ are
compensated by those which come from the corresponding redefinition of 
Wilson's coefficients, even in the general 
case of matricial mixing).
All we need at this point is to look for combinations in which 
 the non universal factors cancel
among each other as in the case of standard amplitude ratios. This can be
performed in two steps. First by combining the correlators in appropriate ratios
to eliminate
the trivial overall dependence on $K_j, K_k$
\footnote{Depending on the particular case under study, it
may be simpler to eliminate this dependence by using  one point functions.
 However in this letter,
 following the discussion at the beginning of this section, 
we shall assume \textit{ not} to have access to
 these quantities.},
and then by choosing a 
combination of ratios  to have
 a finite non-zero result in the $t\to 0$ limit
 (this step replaces  the knowledge of the correlation length in the
 standard approach). 

There are in general several  ways to construct these combinations, but it is
clear that the most interesting ones are those which involve the
smallest number of different operators and thus have 
greater chances to be observed in experiments or in simulations.
In this letter we propose two options which --in our opinion--  are the best
candidates. Let us look at them in detail.

\subsection{QFT's with three or more relevant operators.}
Let us choose a model (like for instance the tricritical Ising model)
 which contains at least three non trivial relevant operators (it
doesn't matter if one of the three is the same $\phi_i$ which we use
as a perturbation). Let us call them $\phi_1$,$\phi_2$ and $\phi_3$, and let us construct the
following combinations:
\eq\label{unsubtractedRatios}
R_{1 2}^{i}\equiv \frac{\lan \phi_1 \phi_1 \ran_i\lan \phi_2 \phi_2 \ran_i}
{[\lan \phi_1 \phi_2 \ran_i]^2}
=
\frac{F_{1 1}^{i}F_{2 2}^{i}}{[F_{1 2}^{i}]^2}
\en
and similarly for the pairs ($\phi_1,\phi_3$) and ($\phi_2,\phi_3$).
The small $t$ expansion of the $R_{j k}^{i}$ can be 
reconstructed from that of their components and one finds in general
\eq
R_{j k}^{i}\sim r^{i}_{j k} t^{z_{j k}^{i}}+\cdots
\en
where $r^{i}_{j k}$ and $z_{j k}^{i}$ are simple combinations of the $f_{j k a}^{i}$ and $z_{j k a}^{i}$
which appear in the small $t$ expansions (see Eq.(\ref{eq1})) 
of the correlators involved in the ratio. 

The reader can easily check that the overall metric factors $K_j,K_k$ disappear in each of 
the combinations $R_{j k}^{i}$. 
At this point one can 
construct one or more universal combinations out of these
ratios by  combining the $R_{j k}^{i}$ so as to cancel the leading $t$ dependence.
The $t\to 0$ limit of any one of these
combinations is clearly an universal quantity and can be exactly 
expressed in terms of the Wilson's
coefficients, their derivatives and the one point functions of the theory.  
For instance one such combination could be:
\eq
Q_{1 2 3}^{i}\equiv \lim_{t\to0} 
\frac{(R_{1 2}^{i})^{z_{1 3}^{i}}}{(R_{1 3}^{i})^{z_{1 2}^{i}}} =
\frac{(r_{1 2}^{i})^{z_{1 3}^{i}}}{(r_{1 3}^{i})^{z_{1 2}^{i}}}
\en

\subsection{Subtracted correlators.}
Unfortunately the above construction cannot be applied, for instance, 
to the very interesting
case of the Ising model which has only two non trivial 
relevant operators in the spectrum. However there is a simple modification 
 which allows to overcome this problem.

Let us construct the differences:
\eq
F_{j j,s}^{i}(t)\equiv F_{j j}^{i}(t) -F_{j j}^{i}(0)
\en
which we shall call in the following ``subtracted correlators''. Since 
$F_{j j}^{i}(0)$ is a constant related to the correlator at the critical point,
the subtracted
correlators are almost as easily accessible from a numerical point of view as the
ordinary correlators. With the help of these subtracted correlators we can
construct new combinations, like for instance
\eq\label{subtractedRatios}
R_{j,s}^{i}\equiv \frac{\lan \phi_j \phi_j \ran_{i,s}}
{\lan \phi_j \phi_j \ran_{i}}\hskip 1cm (j=1,2)
\en
which can be suitably combined among them or with $R_{1 2}^{i}$ so as to eliminate
the residual $t$ dependence 
in the $t\to 0$ limit, thus leading to a universal
combination.

\subsection{QFT estimates.}
Let us finally address the question of the theoretical estimate of these new
universal amplitude ratios. In the case of three dimensional $O(N)$ scalar models
 it should be possible to   obtain
$\epsilon$-expansion (or fixed dimensional) 
estimates by following the same lines used to describe
short distance behavior of field field correlator (see~\cite{PVReport}). 
The
situation is much better in
 two dimensions where, thanks to the progress of CFT
 one knows exactly the operator content, scaling dimensions  and correlators of the
 underlying critical QFT. 
Moreover IRS method gives an exact all order integral representation for the  derivatives
of Wilson's coefficients 
$\wh{\partial^n C_{jk}^\alpha/\partial g_i^n}$
in terms of CFT correlators from which accurate or even exact predictions can be obtained. 
For what concerns VEV amplitudes, it is remarkable that the  approximate  technique known as Truncated Conformal Space 
allows for theoretical predictions of $A_\alpha^{i}$ (see \cite{TCS} and references therein)
and  
in addition in  the case of 2d integrable perturbations many $A_\alpha^{i}$ can be 
evaluated exactly~\cite{fz, Fateev:1998xb}.
Notice that in 2d,
integrable perturbations exist that are not described in terms of 
$O(N)$ models. 

\section{Example: the magnetic perturbation of the Ising model}
The Ising model perturbed by the spin operator $\sigma$ (in $d=2,3$)
is the simplest non-trivial example of the previous discussion.
The perturbing parameter in this case is the magnetic field $h$
(to simplify notations we will drop in this section  
the perturbation label from amplitudes and ratios). 
The 
vacuum expectation values of the two relevant operators are:
\eqa
&&
\lan \sigma \ran_h =  A_\sigma~ 
h^{\frac{2  \Delta_\sigma}{d-2 \Delta_\sigma}} ;
 \;\;\;
\lan \epsilon  \ran_h =  A_\epsilon~
h^{\frac{2 \Delta_\epsilon }{d-2 \Delta_\sigma}}
\ena
In the following we will concentrate on the construction of our new short distance amplitude ratios;
for different amplitude ratios  testing other
regimes of the model, see ~\cite{Delfino:1997ck,Korff:2002ej}.
The first few terms of 
the short distance expansions for the correlators look like:
\eqa
F_{\sigma \sigma}(t)&\sim&
 \wh{ C_{\sigma \sigma}^{\mathbf 1}} + 
 A_\epsilon~\wh{C_{\sigma \sigma}^{\epsilon}} t^{\frac{2 \Delta_\epsilon }{d-2 \Delta_\sigma}} +
 A_\sigma  ~\wh{\dep_h C_{\sigma \sigma}^{\sigma}} t^{1+\frac{2  \Delta_\sigma}{d-2 \Delta_\sigma}}
\nonu 
\\
F_{\epsilon \epsilon}(t) &\sim&
\wh{C_{\epsilon \epsilon}^{\mathbf 1}} + A_\epsilon\wh{C_{\epsilon \epsilon}^{\epsilon}}t^{\frac{2 \Delta_\epsilon }{d-2 \Delta_\sigma}}+
A_\sigma  ~\wh{\dep_h C_{\epsilon \epsilon}^{\sigma}} t^{1+\frac{2  \Delta_\sigma}{d-2 \Delta_\sigma}} \nonu
\\
F_{\sigma \epsilon}(t) &\sim&
A_\sigma  ~\wh{C_{\sigma \epsilon}^{\sigma}} t^{\frac{2  \Delta_\sigma}{d-2 \Delta_\sigma}} +
\wh{\dep_h C_{\sigma\epsilon }^{\mathbf 1}}~ t 
\nonu
\ena
where $t\equiv h r^{d-2\Delta_\sigma}$, $\wh{\dep^n_h C_{ij}^k}$ are the 
rescaled 
derivatives w.r.t. $h$ of Wilson's coefficients.
We warn the reader that the ordering of powers of $t$ from analogous contributions
 is in general different in $d=3$ and $d=2$ due to different scale dimensions. 
The exact values of the first terms of  short distance expansion 
 for  the case $d=2$ are discussed
in~\cite{gm2,cgm00}; here we only need to know that 
in $d=2$
the coefficients $\wh{C_{\epsilon \epsilon}^{\epsilon}}$ 
and $\wh{\dep_h C_{\epsilon \epsilon}^{\sigma}}$ are exactly zero 
and 
$F_{\epsilon \epsilon, s, d=2}=
\frac{1}{2} \wh{\dep_h^2 C_{\epsilon \epsilon}^{\mathbf 1}} t^2+O(t^{32/15})
$ (see \cite{cgm00}).
Let us now introduce the correlator ratios as in  Eq.(\ref{unsubtractedRatios}) 
and Eq.(\ref{subtractedRatios}). 
(Note that leading behavior  of subtracted operators turns out to be dimension dependent.)
\eqa
&&R_{\sigma\epsilon}(t)\equiv\frac{F_{\sigma \sigma}(t) ~F_{\epsilon \epsilon}(t)}
{F_{\sigma \epsilon}^2(t)}\sim
\frac{\wh{C_{\sigma \sigma}^{\mathbf 1}} ~\wh{C_{\epsilon \epsilon}^{\mathbf 1}}}
     {(\wh{C_{\sigma \epsilon}^{\sigma}}~A_\sigma)^2}
~t^{-{\frac{4 \Delta_\sigma}{d-2 \Delta_\sigma}}} +\cdots 
\nonu\\
 && R_{\sigma,s}(t)\equiv
 \frac{F_{\sigma \sigma,s}(t)}{F_{\sigma \sigma}(t)}\sim
A_\epsilon~\frac{\wh{C_{\sigma \sigma}^{\epsilon}}}{\wh{C_{\sigma \sigma}^{\bf 1}}}
t^{\frac{2 \Delta_\epsilon }{d-2 \Delta_\sigma}} +\cdots
\nonu\\
&&R_{\epsilon,s,d=2}(t)\equiv
 \frac{F_{\epsilon \epsilon,s}(t)}{F_{\epsilon \epsilon}(t)}\sim
\frac{1}{2} ~\frac{\dep_h^2 \wh{C_{\epsilon \epsilon}^{{\bf 1} } }} 
{\wh{C_{\epsilon \epsilon}^{\bf 1}}} t^{2} + \cdots 
\nonu \\
&&R_{\epsilon,s,d=3}(t)\equiv
 \frac{F_{\epsilon \epsilon,s}(t)}{F_{\epsilon \epsilon}(t)}\sim 
A_\epsilon
\frac{ \wh{C_{\epsilon \epsilon}^{\epsilon}}}
 {\wh{C_{\epsilon \epsilon}^{\bf 1}}} t^{\frac{2 \Delta_\epsilon }{3-2 \Delta_\sigma}} +\cdots 
\nonu\ena
 It is easy to see that the simplest combinations of these  ratios 
 which in the $t\to0$ limit lead to  universal quantities are
 \eqa
 Q_{\sigma\epsilon,s}(t)&\equiv&
 R_{\sigma\epsilon}^{\frac{\Delta_\epsilon}{2\Delta_\sigma}}(t)R_{\sigma,s}(t)\\
 Q^{'}_{\sigma\epsilon,s,d=2}(t)&\equiv& 
  R_{\epsilon,s,d=2}^{-\frac{4}{15}}(t)R_{\sigma,s}(t)\\
 Q^{'}_{\sigma\epsilon,s,d=3}(t)&\equiv& R_{\epsilon,s,d=3}^{-1}(t)R_{\sigma,s}(t)
\ena
$Q^{'}_{\sigma\epsilon,s}$ is probably the best candidate in $d=3$, while in 
 $d=2$ the best choice is to study $Q_{\sigma\epsilon,s}$. 
Let us concentrate on $d=2$ case, where $\frac{\Delta_\epsilon}{2\Delta_\sigma}=4$,
 and the
 universal ratio can be written directly in terms of correlators as:
 \eq
 Q_{\sigma\epsilon,s,d=2}(t)=
 \frac{\lan \sigma\sigma \ran_{h,s}
  (\lan \sigma\sigma \ran_h)^3 
(\lan \epsilon\epsilon \ran_h)^4}
{(\lan \sigma\epsilon \ran_h)^8}.
\en

The ratio $ Q_{\sigma\epsilon,s,d=2}(0)$ can be obtained exactly.
If we choose the CFT field normalizations such
$\wh{C_{\sigma \sigma}^{\bf 1}}=\wh{C_{\epsilon \epsilon}^{\bf 1}}=1$ we have
$\wh{C_{\sigma \epsilon}^{\bf \sigma}}=
\wh{C_{\sigma \sigma}^{\bf \epsilon}}=\frac12$. Thus we find:
\eq
Q_{\sigma\epsilon,s,d=2}(0)=128\frac{A_\epsilon}{{A_\sigma}^8}
\en
By using the exact values of the VEV amplitudes 
(see ~\cite{fz,Fateev:1998xb} and references therein)
\eq
A_\sigma=1.27758227...  \hskip 2cm
A_\epsilon=2.00314...
\en
 we finally obtain
$Q_{\sigma\epsilon,s,d=2}(0)=36.125...$.

\vskip .5 cm
{\bf  Acknowledgements:}
We thank   A.~Pelissetto and J.~Zinn--Justin for helpful 
suggestions. 
This work was partially supported by the 
European Commission TMR program HPRN-CT-2002-00325 (EUCLID).
The work of one of us (P.G.) is supported by the COFIN "Teoria dei 
campi, meccanica statistica e sistemi elettronici".

\end{document}